\documentclass[12pt]{article}
\usepackage{a4wide}
\newcommand{\be}{\begin{equation}}
\newcommand{\ee}{\end{equation}}
\newcommand{\ba}{\begin{eqnarray}}
\newcommand{\ea}{\end{eqnarray}}
\begin{document}
\begin{flushright}
LU TP 01-37\\
December 2001
\end{flushright}
\begin{center}
{\Large\bf Comment on the Pion Pole Part of the
Light-by-Light Contribution to the Muon $g-2$}
\\[1cm]
{\bf Johan Bijnens}\\Department of Theoretical Physics 2,
 University of Lund\\
S\"olvegatan 14A, S 223 62 Lund, Sweden\\[0.25cm]
{\bf Elisabetta Pallante}\\
S.I.S.S.A., Via Beirut 2-4, 34014 Trieste, Italy
\\[0.25cm]
{\bf Joaquim Prades}\\
Centro Andaluz de F\'{\i}sica de las Part\'{\i}culas
Elementales (CAFPE) and\\
Departamento de F\'{\i}sica Te\'orica y del Cosmos, Universidad 
de Granada, \\
Campus de Fuente Nueva, E-18002 Granada, Spain 
\end{center}
\begin{abstract}
We comment on the recent calculations of the pion pole part
 of the light-by-light contribution to the muon anomalous magnetic moment and
 we point out where the analysis in our previous work was mistaken. 
\end{abstract}

In the recent paper \cite{KN01}, 
the pion pole part of the light-by-light contribution to the muon
anomalous magnetic moment was reevaluated 
and  analytical expressions for a large class
of $\pi^0\gamma^*\gamma^*$ form factors fulfilling 
OPE and large--$N_c$ QCD constraints were obtained.
In \cite{KNPR01}
the leading double logarithmic dependence in the case of point-like
pion-photon-photon couplings was derived analytically.
This result has been confirmed  in \cite{BCM01}.
The absolute value of the contribution was found to be in good
agreement with the evaluations done previously 
\cite{BPP95,BPP96,HKS96,BAR01,BP01}, but opposite in sign.

We have gone back to our old programs and notes in order to find 
the source of the discrepancy. 
The pion pole contribution, as well as all the others,
was calculated twice independently inside our collaboration and afterwards the
results were compared. Both analyses agreed and had
the correct sign analytically, but unfortunately
both analyses made the same sign mistake in the overall normalization factor
in putting it into the integration program.
Since both computations agreed and also agreed
with previously known results we saw no reason to doubt our results
for the pion pole.
More studies of variations in the form factor \cite{BP01} did also not 
indicate that there was a major problem with the size of the form-factor.
The same sign problem also
affects the axial-vector contribution, but not the others we discussed.
The scalar exchange contribution in our approach is linked by chiral symmetry
constraints to the quark-loop low-energy contribution and had the correct sign.

We have also checked that the analytical
formulas for the amplitudes  in \cite{BPP96} are all correct with the 
exception of an overall minus sign missing in Eq. (2.7), 
that was an errata in the manuscript and not the origin of the sign mistake
in the pion pole contribution.

The often quoted argument that there
was no matching with the short-distance quark-loop contribution
was tested in our work, in particular in Section 7.1 of 
\cite{BPP96}.
There, it was found that the short-distance quark loop matched 
perfectly well numerically
when added together with another contribution that our 
treatment of the low-energy part required. For us, there was thus no 
reason to doubt that we had a reasonable matching with the expected 
short-distance behaviour even with the pion-pole and the quark-loop
contributions having a different sign.

One should also not forget that, while the pion and the other 
pseudoscalar pole contributions are the dominant ones, there 
are other contributions
at the 20 to 30 percent level that also need to be estimated. In the
works \cite{BPP96,HKS96} 
they have been evaluated and were found to cancel
to a large extent. We intend to reevaluate all contributions and come
back to them in a later more extended note.
Our number corrected for the sign mistakes becomes
\ba
a_\mu^{\rm LbL}&=& (2.1\pm0.3)\cdot 10^{-10}[\mbox{quark-loop}]
+(-0.68\pm0.2)\cdot 10^{-10}[\mbox{scalar}]
 \nonumber \\ &+&(8.5\pm1.3)\cdot 10^{-10}[\mbox{pseudoscalar}]+
(0.25\pm0.1)\cdot 10^{-10}[\mbox{axial-vector}]
\nonumber\\&+& (-1.9\pm1.3)\cdot 10^{-10}[\pi K\mbox{-loop}]
\nonumber\\&=& (8.3\pm3.2)\cdot 10^{-10}.
\ea
We have also checked that using the pointlike coupling our cut-off dependence
numerically agrees with the one calculated analytically in \cite{KNPR01}.

A similar note \cite{HH01} has appeared by the authors of the other full
evaluation \cite{HKS96}.

\end{document}